\documentclass[prb,preprint,aps,superscriptaddress,longbibliography]{revtex4-1}
\usepackage[]{color}
\usepackage{appendix}
\usepackage{graphicx, epstopdf}
\usepackage{subfigure}
\usepackage{float}
\usepackage{amsmath}
\usepackage{amsfonts}
\usepackage{amssymb}
\usepackage{bm}% bold math
\usepackage{subfigure}
\usepackage{caption}
\usepackage{hyperref}
\usepackage{xr}
\usepackage{lineno}
\usepackage{natbib}
\usepackage{soul}
\usepackage{amssymb}
\usepackage{multirow}
\usepackage[T1]{fontenc}
\usepackage{booktabs}
\usepackage{lipsum}
\usepackage{mathrsfs}
\usepackage{xr}
\begin{document}
\makeatletter
\newcommand{\rmnum}[1]{\romannumeral #1}
\newcommand{\Rmnum}[1]{\expandafter\@slowromancap\romannumeral #1@}
\makeatother

\title{Interpretable machine learning of formation energies and band gaps in hybrid perovskites through SHAP analysis and SISSO modelling}

%\title{\textbf{Perovskite Research Paper}}
%\date{}
\author{Christopher Xia}
\affiliation{Saint Francis High School, Mountain View, CA}

\author{H. Zhang}	
\affiliation{College of Future Information Technology, Fudan University, Shanghai 200433, China}

\author{Q. Y. Jin}
%\email{qyjin@phy.ecnu.edu.cn}
\affiliation{School of Physics, East China Normal University, Shanghai 200062, China}

\begin{abstract}
%Predicting formation energies and band gaps across large perovskite chemical spaces requires accurate and interpretable machine-learning models. Here, we combine XGBoost and graph neural networks to screen hybrid organic–inorganic perovskites and examine the features underlying their predicted properties. Trained on 4,456 compounds, the optimized XGBoost models achieved high $R^2$ values for both formation energy and band gap. Screening over 180,000 candidates for negative formation energies and predicted band gaps between 0.5 and 2.0 eV yielded 16,937 compounds for further evaluation. Shapley additive explanations (SHAP) analysis identified influential molecular, elemental and geometric features, while correlation analysis reduced input redundancy. The retained primary features were subsequently used in sure independence screening and sparsifying operator (SISSO) modelling to derive four-dimensional analytical expressions for each property. The formation-energy expression captures coupled contributions from molecular electronic properties and B- and X-site elemental characteristics, whereas the band-gap expression emphasizes relative atomic sizes and B-site properties. This work combines efficient screening with compact representations of learned structure–property dependencies, providing a basis for prioritizing and interpreting promising perovskite candidates for subsequent first-principles calculations and experimental validation.
Accurate and interpretable property prediction is essential for exploring hybrid organic–inorganic perovskites. Here, we combine machine-learning screening with SHAP analysis and SISSO descriptor discovery. Using 4,456 compounds, optimized XGBoost models achieved test-set $R^2$ values of 0.99 for formation energy and 0.93 for band gap and were applied to over 180,000 candidates. Comparison across approximately 17,000 common compounds revealed broadly overlapping  distributions from XGBoost and CGCNN but substantially lower predictions from UPET, suggesting potential limitations in its transferability to hybrid organic–inorganic perovskites. SHAP analysis and correlation-based pruning retained eight primary features for each property, from which SISSO derived expressions comprising four nonlinear descriptors. These expressions couple molecular and inorganic properties for formation energy and emphasize relative atomic sizes and B-site characteristics for band gap. This framework supports efficient candidate prioritization and interpretable analysis of learned structures–property relationships in hybrid perovskites.
\end{abstract}

\maketitle

\section{Introduction}

As climate change continues to worsen, the demand for renewable energy is on the rise. In particular, solar energy is very abundant, leading to photovoltaics becoming one of the renewable energy sources with the most potential. Currently, the photovoltaic market is dominated by silicon based solar cells. Although they currently present a very respectable power-conversion efficiency (PCE), they also struggle with massive upfront production energy requirements and methods of quality recycling\cite{dubey2013socio}. 
%
%Fortunately, perovskite based solar cells have been at the front of solar cell research. Not only are they relatively cheap and easy to manufacture, they have already shown PCE that can rival silicon based solar cells, and have many properties that give it massive potential for further optimization \cite{miah2024band}. However, there are still concerns plaguing perovskite solar cells, primarily their stability and the prevalent use of toxic $Pb^{2+}$, which is especially prone to leeching into the environment\cite{li2020biological}. As a result, searching for new, promising materials is crucial for the long term success of completely environmentally friendly perovskite solar cells. 
Among these alternatives, hybrid organic–inorganic perovskites have attracted considerable attention owing to their rapidly improving PCE\cite{kojima2009organometal}, potentially low-cost fabrication and tunable optoelectronic properties.

The photovoltaic potential of these materials was established through early demonstrations of perovskite sensitizers, solid-state devices and sequential deposition methods\cite{kojima2009organometal,lee2012efficient,burschka2013sequential,miah2024band}. Their molecular and inorganic constituents provide a broad compositional space for tailoring electronic properties and structural stability. Subsequent compositional engineering demonstrated that mixed-cation formulations can improve phase stability and device performance\cite{jeon2015compositional,saliba2016cesium}. However, material degradation and the environmental risks associated with lead-containing compositions, including the potential release of lead upon degradation, remain major barriers to widespread deployment\cite{li2020biological}. Addressing these challenges requires identifying alternative compositions that combine suitable band gaps with favourable energetics and reduced toxicity. The breadth of the available chemical space therefore motivates efficient screening strategies that can prioritize promising candidates and reveal the compositional relationships governing their properties.

Double perovskites expand this search space by accommodating distinct B-site cations. Experimental and computational studies of silver–bismuth and silver–indium halide double perovskites demonstrated lead-free substitution routes and substantial variation in their electronic properties\cite{slavney2016bismuth,mcclure2016cs2agbi,volonakis2016leadfree,volonakis2017cs2inagcl6}. These inorganic systems illustrate the opportunities offered by B-site chemistry, while hybrid double perovskites introduce additional molecular degrees of freedom\cite{cai2022discovery}. The resulting diversity creates a large combinatorial space that is difficult to explore exhaustively. Identifying alternative compositions therefore requires balancing electronic properties and energetic favourability. %Screening strategies must therefore identify promising candidates efficiently while resolving the compositional dependencies associated with their predicted properties.

Formation energy and band gap provide complementary screening criteria. Formation energy measures energetic favourability relative to specified reference states, but does not alone establish resistance to decomposition into competing phases\cite{nagabhushana2016direct}. Geometric compatibility, including the size relationships underlying tolerance-factor descriptions, also influences perovskite structure formation\cite{travis2016tolerance}. Molecular steric effects can modify octahedral tilting and metal–halide–metal bond angles, changing orbital overlap and band gaps\cite{filip2014steric}. For representative Pb- and Sn-based hybrid halide perovskites, electronic-structure calculations further demonstrate the importance of spin–orbit coupling and relativistic effects\cite{even2013spinorbit,umari2014relativistic}. These coupled structural and electronic contributions motivate the use of molecular, elemental and geometric descriptors in predictive modelling.

\textit{ab-initio} calculations are essential for evaluating such properties, but become computationally demanding across large candidate sets. Machine learning (ML) provides efficient surrogate models for prioritizing subsequent calculations. Composition-based frameworks and element-wize descriptors have enabled predictions of multiple materials properties\cite{ward2016general,isayev2017universal}. Pilania et al. demonstrated descriptor-based band-gap prediction for double perovskite oxides\cite{pilania2016bandgaps}, while Cai et al. combined machine learning and high-throughput calculations to explore hybrid double perovskites\cite{cai2022discovery,cai2023discovery}. %For tabulated features, XGBoost offers a scalable approach to learning nonlinear dependencies.[20] 
However, compositional descriptors may only indirectly capture local coordination and structural distortions.
Graph neural networks (GNN) provide complementary representations of atomic environments. CGCNN learns properties from crystal graphs\cite{xie2018cgcnn}, whereas MEGNet and ALIGNN demonstrate further approaches to incorporating structural information, including explicit bond-angle information in ALIGNN\cite{chen2019graph,choudhary2021alignn,lee2026alignn}. More recently, PET-MAD-DOS, accessible through the universal PET (UPET) framework, has enabled electronic density-of-states prediction and band-gap estimation across diverse organic and inorganic systems\cite{how2026universal}. Broad training coverage nevertheless does not guarantee uniform accuracy for every material family. %Comparisons on common compounds can reveal differences associated with model representations and training conditions, although agreement between prediction distributions cannot independently establish accuracy.

Interpretability provides a complementary basis for evaluating these predictions. SHAP attributes feature contributions to individual predictions, with tree-based implementations also supporting the analysis of global model behaviour\cite{lundberg2017unified,lundberg2020local,cai2022discovery}. Descriptor selection is itself central to reliable materials modelling\cite{ghiringhelli2015descriptor}. The Sure Independence Screening and Sparsifying Operator (SISSO) and its C++ implementation, SISSO++, identify sparse analytical relationships by searching nonlinear combinations of primary features\cite{ouyang2018sisso,purcell2022sisso++}. The SISSO-derived tolerance factor of Bartel et al. illustrates its ability to construct compact descriptors for perovskite formability\cite{bartel2019tolerance,yang2026octahedral}. Combining SHAP-guided selection with correlation-based pruning can reduce input redundancy and restrict the SISSO search space, linking feature attribution to explicit empirical expressions. %Such expressions make the coupled dependencies easier to inspect, but remain empirical models whose coefficients and descriptors must be interpreted within the sampled chemical space and the chosen feature definitions.

In this work, we combine predictive modelling, model comparison and descriptor discovery for hybrid perovskites. Regression models are assessed using 4,456 compounds, and optimized XGBoost models are applied to 180,057 candidates\cite{chen2016xgboost}. Band-gap predictions from XGBoost, CGCNN and UPET are compared across 17,001 common compounds. SHAP analysis and correlation-based pruning select eight primary features for each property, from which SISSO derives expressions comprising four nonlinear descriptors. These expressions relate molecular electronic properties, relative atomic sizes and B- and X-site characteristics to the learned formation-energy and band-gap trends, supporting efficient candidate prioritization and interpretable analysis.

\section{Data preparation and Methodology}

The training data set for bandgap and formation energy consists of 4456 double hybrid organic-inorganic perovskite (HOIP) crystal compounds and has the same physical properties presented in the candidate dataset\cite{cai2022discovery}, but with the inclusion of the band gap and formation energy as well in each respective data set\cite{cai2022discovery}. The training dataset was first imputed to replace all NaN’s with 0’s, then went through dimensionality reduction with PCA method\cite{pedregosa2011scikit}, then underwent an 80-20 train test split before training the regression models. The prediction dataset contains over 180,000 HOIP crystal candidates\cite{cai2022discovery}. Many physical properties of the individual ions and bonds in each candidate were noted, including ionic radii, the Mendeleev number, dipole polarizability, etc. These features are used primarily used in the training and inference of the regression models, but the chemical composition of the crystals is used through every other step of screening.

%\subsection{General Workflow}

%Generally, discovery of new perovskite solar cell materials requires high throughput \textit{ab initio} calculations or experiments. Although \textit{ab initio} calculations prove to be a highly accurate high throughput method, they are still fairly computationally expensive, especially when it is applied to crystal materials with large unit cells. Thus, recently machine-learning/deep learning methods have been developing to predict electronic-structures and enable efficient simulations of large-scale material systems. In this work, we focus on filtering promising electrical and physical properties, i.e. suitable bandgaps and formation energies. %The framework of our work composes of three layers of filtering, with each layer becoming more accurate and more computationally expensive. The screening begins with trained machine-learning regression models for formation energies and bandgaps, respectively, followed by a recalculation from GNN models. 
%Finally, structural relaxation is performed to remove artificial strain within the compounds to prepare for further analysis. In this work, the focus was on finding materials with promising electrical and physical properties -- namely, a suitable bandgap and formation energy.
The framework of our work consists of three layers of screening, with each successive layer providing higher accuracy at the cost of increased computational expense. The screening process begins with trained machine-learning regression models for predicting formation energies and band gaps, respectively, followed by more accurate recalculations using graph neural network (GNN) models. Finally, the SISSO method is employed to derive empirical formula for the formation energies and band gaps of perovskite materials, providing physical insights into the underlying structure-property relationships, and further enabling more efficient and interpretable materials screening.

Regression models are computationally inexpensive. However, their accuracy is somewhat lacking compared to other methods. As a result, they are very effective for the first step of filtering out the majority of candidates that are either invalid or do not have a good bandgap. The intention was to train two regression models, one for predicting the bandgap and one for predicting the formation energy of the compound. Using the two regression models, we performed an initial screening to find materials that showed promising bandgap and formation energy.

%\subsection{Deep Learning Models}

Although the regressions prove to be a useful form of initial screening, they are simple regression models and do not consider the structures of the perovksites as a feature for its predictions. %However, there are still too many remaining candidates to perform first principle calculations efficiently. 
Therefore, we used two deep learning models, i.e. CGCNN\cite{xie2018cgcnn} and UPET\cite{PET-MAD-DOS-2025}, as an intermediate step to recalculate the bandgap with greater accuracy than the regression model but still at a faster rate than first principle calculations. These models also take in the structure of the HOIP crystals to make a more accurate prediction of both formation energy and bandgap. %In the end, we used both models simultaneously and filtered out the materials that both models agreed on having a good bandgap ($1.0$ eV $< E_g <$ $2.0$ eV) to move on to final verification.
%
%\subsubsection{CGCNN}
%Crystal Graph Convolutional Neural Networks, or CGCNN, is a deep learning model that predicts material properties by using the crystal structure as the only input\cite{Xie_2018}. 
We trained a CGCNN model on 3489 organic-inorganic perovskite materials with their crystal structures and bandgaps, then fed the model with the crystal files of the screened candidates into the neural network\cite{cai2022discovery}. The crystal files for our candidates were made by taking a template perovskite and swapping the elements in the A, B, and X positions, which allows for a fast way to generate these crystal structure files. 
%
%\subsubsection{UPET}
Simultaneously, %we used a Point Edge Transformer (PET) graph neural network for this level of the filtering. Specifically, 
we used PET-MAD-DOS\cite{PET-MAD-DOS-2025}, to predict bandgaps of generated HOIPs. Although it is not specifically trained on perovskite crystals, it is trained on the Massive Atomic Diversity Dataset, creating a robust model that would be a good reference to compare alongside CGCNN.

% \subsection{SISSO}

% Sure Independence Screening and Sparsifying Operator, or SISSO is an algorithm that takes different features and combines them using a set of operators (+, -, $\times$, $\div$, $x^{-1}$, $x^2$, \sqrt{x}, $e^x$) in order to predict another target property. 

For the interpretable descriptor-discovery stage, the SISSO method was employed to establish compact analytical relationships between the input features and target material properties\cite{ouyang2018sisso,purcell2022sisso++}. Starting from the primary descriptors, SISSO generates a large feature space by recursively applying a predefined set of mathematical operators, including \(+\), \(-\), \(\times\), \(\div\), \(x^{-1}\), \(x^2\), \(\sqrt{x}\), and \(e^x\). Sure independence screening is then used to reduce the dimensionality of the generated feature space, followed by the sparsifying operator to identify a small subset of optimal descriptors for property prediction. %This approach provides low-dimensional, physically interpretable expressions that complement the predictions obtained from machine-learning models.
%
%For our best performing model inputs, it had a descriptor dimension of 3 and an fcomplexity of 2. 
We set the minimum and maximum values of each feature to be $10^{-3}$ to $10^5$ before it is discarded. Furthermore, we allowed a nonzero intercept, used the L0 sparse regression method, and used RMSE to grade each dimension. Once the model has been returned, we predicted the bandgaps of 180,000 candidates and compared them to the predictions of XGBoost through correlation analysis.

\section{Results and discussion}

% \subsection{SHAP Analysis and Feature Importance}

% After selecting our model, we conducted a SHAP analysis to find the most important features for each regression model. We also generated correlation plots for the 10 most important features as well. 
\subsection{Machine/Deep-learning models for formation energies and bandgaps}

There were multiple types of regression algorithms tested for formation energy: Gradient Boosting Regressor (GBR), Kernel Ridge Regressor (KRR), Random Forest Regressor (RFR), Decision Tree Regressor (DTR), K Neighbors Regressor (KNR), Support Vector Regressor (SVR), Multi-Layer Perceptron Regressor (MLP), Extra Trees Regressor (ETR), and XGBoost (XGB) methods\cite{pedregosa2011scikit}. % The training dataset contains 4456 crystals with crystal formula, properties, as well as real formation energy. The dataset was first imputed to replace all NaN’s with 0’s, then went through dimensionality reduction with PCA,then underwent an 80-20 train test split before training the regression models.

\begin{figure}[ht!]
    \centering
    \includegraphics[width=1\textwidth]{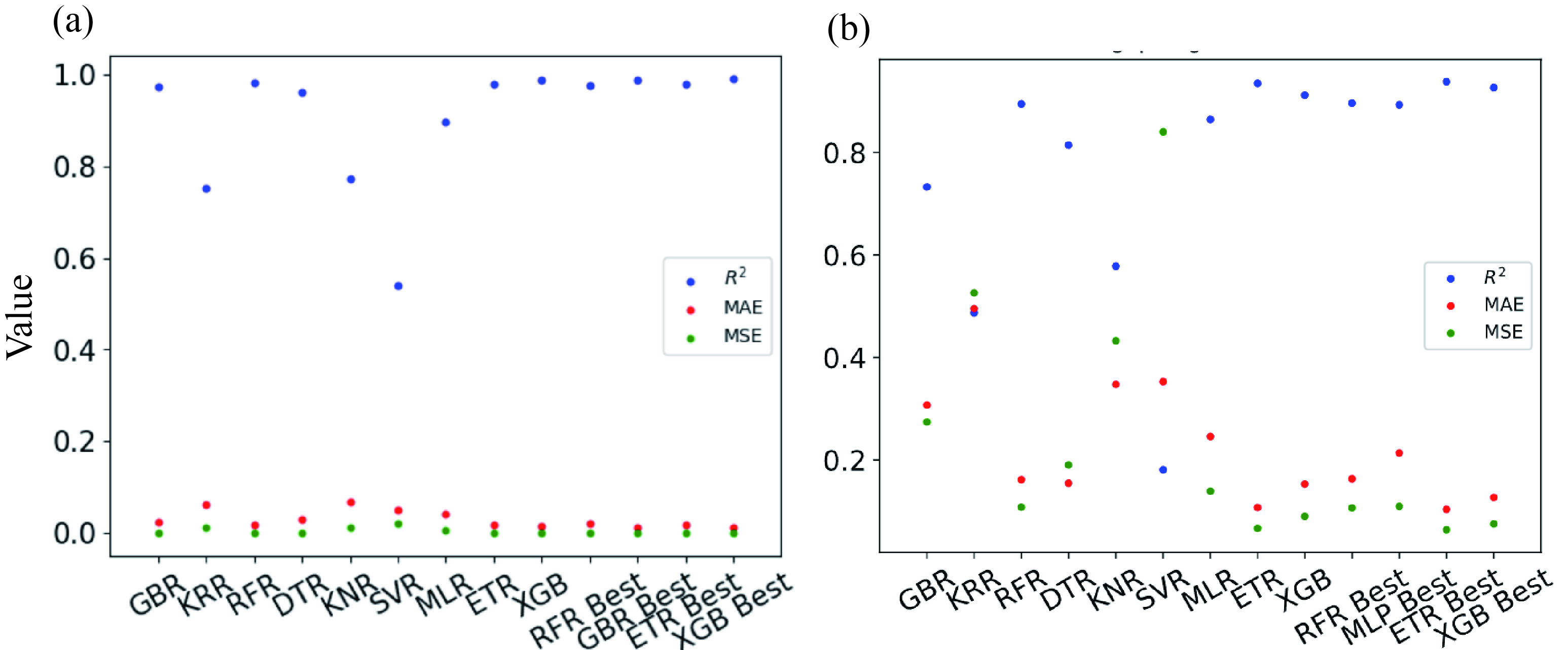}
    \caption{Performance of various models on predicting (a) formation energies and (b) bandgaps of hybrid perovskites. For RFR, GBR, ETR, and XGBoost, hyperparameter optimzation was also conducted.}
    \label{fig:mls}
\end{figure}

Initially, as shown in Figure~\ref{fig:mls}(a), the top performing regression models for formation energies of perovskites include XGBoost ($R^2=0.99$), ETR ($R^2=0.98$), RFR ($R^2=0.98$), and GBR ($R^2=0.97$) . These four models were all also able to undergo hyperparameter optimization. However, the changes in accuracy were marginal, and both ETR and RFR even saw a slight decrease in fit. The biggest increase in $R^2$ was seen in GBR, which increased its $R^2$ by 0.02 to 0.99. After the optimization, the best performing model overall is XGBoost with an $R^2$ of 0.99.

As for bandgap calculations, as shown in Figure~\ref{fig:mls}(b), we compared the same regression models as for calculating formation energy. The models were not as accurate. Initially, the best performing regression models were RFR ($R^2=0.89$), ETR ($R^2=0.94$), and XGBoost ($R^2=0.91$). After performing hyperparameter optimization on XGB, RFR, MLP, and ETR, there was some improvement, but not by much. The best regression model was still ETR ($R^2=0.94$) with XGBoost as a close second ($R^2=0.93$).

As a result, XGBoost was chosen as it was the best formation energy regression model and was very close to being the best bandgap regression model. Then, with one XGBoost regression model trained for bandgaps ($E_g$) and another for formation energies ($E_f$), we predicted bandgaps and formation energies for each of the 180,057 crystals\cite{cai2022discovery}. An initial formation-energy filter ($E_f < -0.1$ eV/atom) was applied to identify energetically favourable candidates, followed by  screening using three machine-learning and deep-learning models.

%Finally, before going onto the next step, we focused the candidates pool even more by isolating candidates with Cs in the B site and removing candidates that contained expensive, toxic, and ineffective elements. As a result, we screened for cesium based perovskites, and was left with 1788 candidates in total for the next layer of filtering.

% \subsection{GNN models for formation energies and bandgaps}

%After the filtering of the regression model was done, the remaining candidates had their bandgaps predicted by both CGCNN and UPET. Each model was fairly quick, being able to screen a few materials per second even when ran on a CPU. This high speed is considerably better than calculating from first principles, as DFT calculations can take hours running on GPU. However, with sufficiently large datasets, it is still better to have an initial screening with a regression model first. 

 \begin{figure}[ht!]
     \centering
     \includegraphics[width=0.8\textwidth]{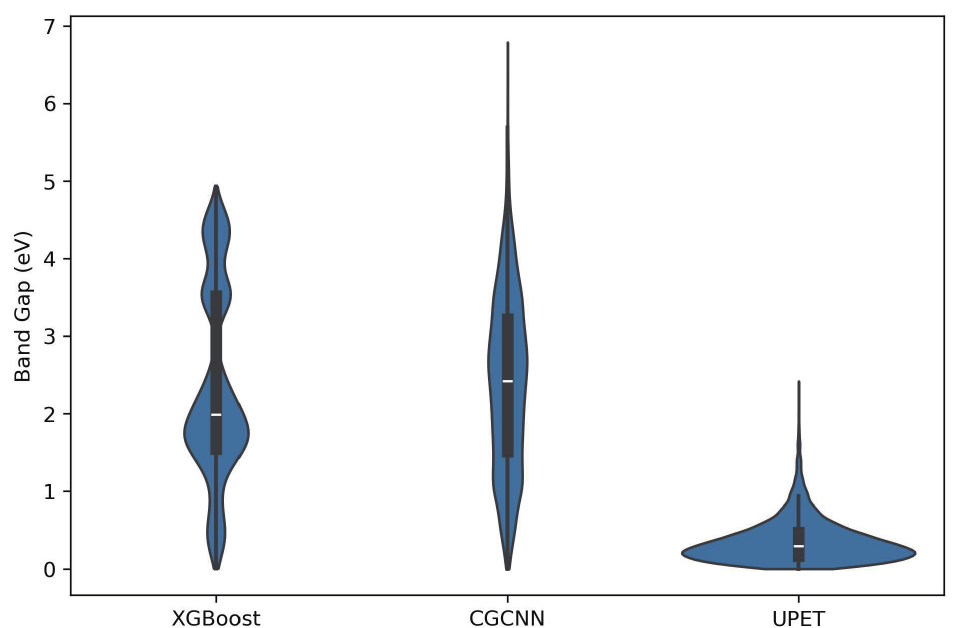}
     \caption{Comparison of band gaps predicted by XGBoost, CGCNN and UPET for a shared dataset containing over 17,000 perovskites.}
     %Comparison of the predicted values of CGCNN, UPET, and XGBoost on all of the materials that pased the linear regression model. }%Although there are a few compounds that all three models agree on, most of the time there is a wide variance.}
     \label{fig:cgcnn_v_upet_v_xgb}
 \end{figure}

Figure~\ref{fig:cgcnn_v_upet_v_xgb} compares the  distributions predicted by XGBoost, CGCNN and UPET for a shared set of 17,001 perovskite compounds. This set was obtained from 18,407 chemical formulas common to all three prediction datasets by excluding compounds with $E_g \leq 0$ eV predicted by any model. The violin widths represent the estimated probability density, the internal boxes indicate the interquartile ranges, and the white markers denote the medians. XGBoost produces a multimodal distribution with a median near 2.0 eV. The CGCNN distribution substantially overlaps with that of XGBoost but has a higher median of approximately 2.4 eV and a longer upper tail. In contrast, UPET predictions are concentrated at much lower energies, with a median near 0.3 eV.

The broadly similar distributions obtained from XGBoost and CGCNN, both trained on perovskite datasets\cite{cai2022discovery}, differ markedly from that predicted by UPET, which was trained on general inorganic crystals\cite{PET-MAD-DOS-2025}. This discrepancy suggests possible limitations in transferring UPET to hybrid organic–inorganic perovskites, where bandgaps depend on the coupling between molecular composition, inorganic coordination and electronic structure. UPET’s lower-energy distribution may reflect insufficient representation of these hybrid environments in its training data, although differences in the electronic-structure methods used to generate training labels could also contribute to the discrepancy. These results highlight the need to validate and, where necessary, calibrate UPET against consistent reference calculations before using it for bandgap screening in this chemical space.

\subsection{SHAP analysis and SISSO-based empirical formula for formation energies and bandgaps}

To elucidate the feature dependencies underlying formation energy and bandgap, we employed the SISSO method to derive compact empirical expressions for both properties. As described above, the most influential features were first identified through SHAP analysis of the XGBoost models, and highly correlated features were removed to reduce redundancy. SISSO then recursively combined the retained features using mathematical operators to construct a large space of candidate descriptors that explicitly capture nonlinear combinations of the input features. Sure independence screening reduced this space to a tractable subset, from which sparse regression identified low-dimensional descriptor combinations for each target property. The resulting expressions relate formation energy and bandgap to a small number of composite descriptors, providing an interpretable representation of the learned structure-property relationships.

%Due to the exponential nature of its search space, we had to greatly limit the features present. Using the SHAP analysis on XGBoost, we were able to isolate the top 10 most important features in the regression model. Then, after generating a correlation matrix, we removed features that were highly correlated with others in the dataset. Out of the features, we removed the octahedral factor due to its high correlation between X/averB and X-averB of -0.96 and -0.95 respectively. X/averB and X-averB are highly correlated to each other as well at 0.95, so we removed X/averB as well. 

\begin{figure}[ht!]
    \centering
    \includegraphics[width=1\textwidth]{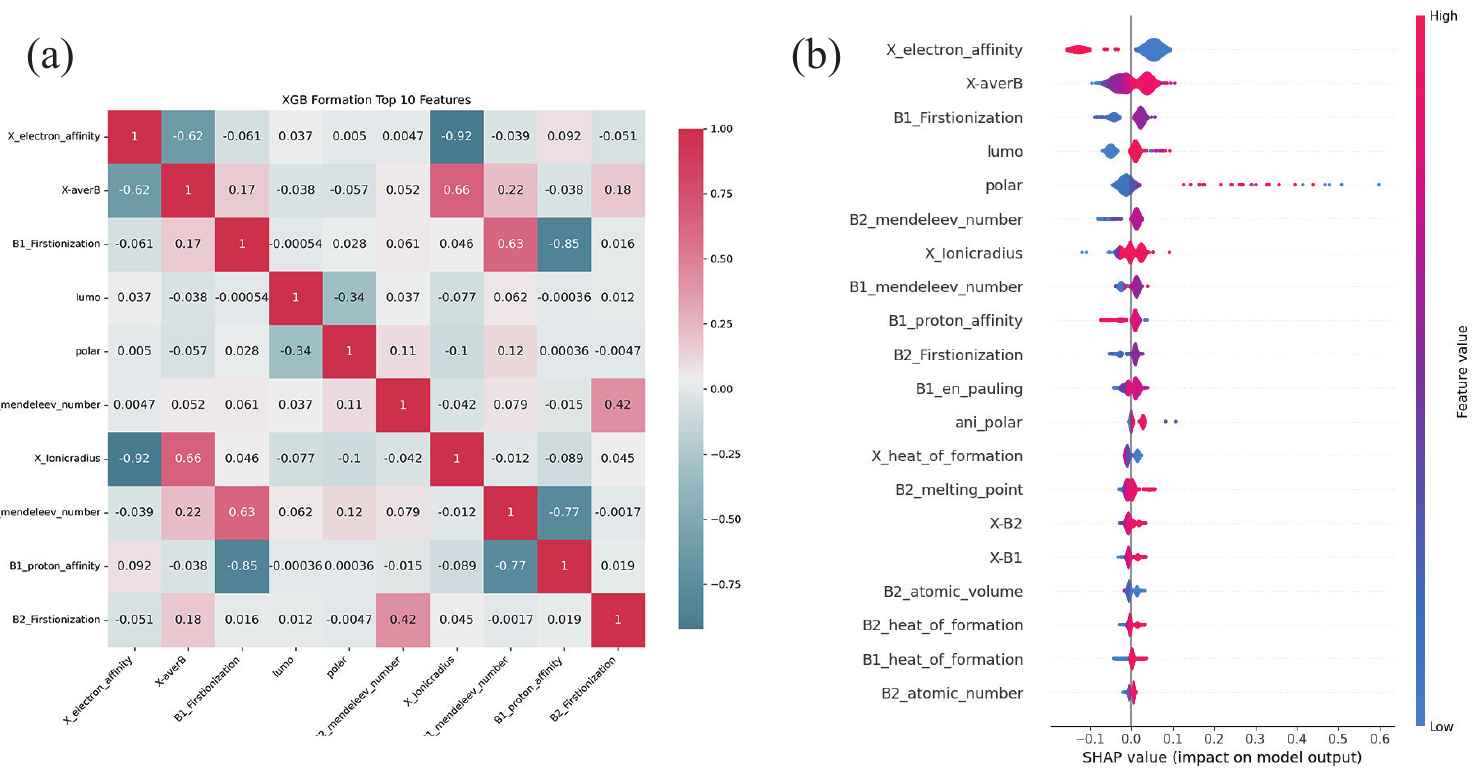}
    \caption{(a) Correlation analysis and (b) SHAP analysis of the most important features influencing the formation energies of perovskites.}
    \label{fig:formation-features}
\end{figure}

Because the SISSO descriptor space grows combinatorially with the number of primary features\cite{ouyang2018sisso}, we restricted the input set to the most influential features identified by SHAP analysis of the XGBoost models. For formation energy as shown in Figure~\ref{fig:formation-features}(b), the ten highest-ranked features in descending order of importance were X$_-$electron$_-$affinity, X-averB, B1$_-$Firstionization, Lumo, Polarizability, B2$_-$mendeleev$_-$number, X$_-$Ionicradius, B1$_-$mendeleev$_-$number, B1$_-$proton$_-$affinity and B2$_-$Firstionization. Pairwise correlations among these features were subsequently examined to assess redundancy, as shown in Figure~\ref{fig:formation-features}(a). The electron affinity and ionic radius of the X-site element were strongly correlated; we therefore retained electron affinity, which ranked higher in the SHAP analysis, and excluded ionic radius. Similarly, the strong correlation between B1$_-$Firstionization and B1$_-$proton$_-$affinity led us to retain the former and exclude the latter. The remaining eight most influential features were retained as inputs for the subsequent SISSO analysis.

\begin{figure}[ht!]
    \centering
    \includegraphics[width=1\textwidth]{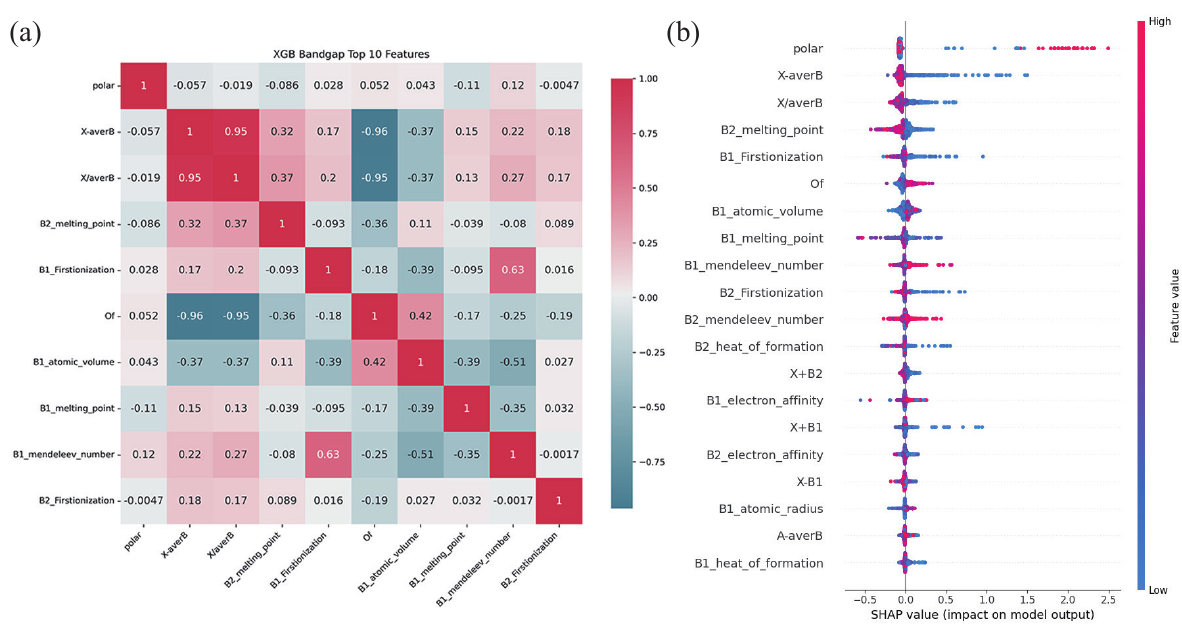}
    \caption{(a) Correlation analysis and (b) SHAP analysis of the most important features influencing the bandgaps of perovskites.}
    \label{fig:bandgaps}
\end{figure}

Similarly, for bandgaps of perovskites as shown in Figure~\ref{fig:bandgaps}(b), the ten highest-ranked features identified by SHAP analysis were Polarizability, X-averB, X/averB, B2$\_$melting$\_$point, B1$\_$Firstionization, Of, B1$\_$atomic$\_$volume, B1$\_$melting$\_$point, B1$\_$mendeleev$\_$number and B2$\_$Firstionization, in descending order of importance. Pairwise correlations among these features were subsequently examined to assess redundancy, as shown in Figure~\ref{fig:bandgaps}(a). The octahedral factor (Of) was strongly correlated with X-averB and X/averB, with correlation coefficients of -0.96 and -0.95, respectively, and was therefore excluded. X-averB and X/averB were also strongly correlated, with a coefficient of 0.95. We retained X-averB, which ranked higher in the SHAP analysis, and excluded X/averB. The remaining eight most influential features were retained as inputs for the subsequent SISSO analysis.

For formation energy, we performed SISSO on a dataset of 4,456 perovskite samples\cite{cai2022discovery}, with a maximum descriptor dimension of four. The selected model expresses formation energy as a linear combination of four nonlinear descriptors constructed from the eight primary features retained after SHAP ranking and correlation-based pruning. At each descriptor dimension, the sure independence screening (SIS) step retained the 100 highest-ranked candidate features (nf$\_$sis = 100) for subsequent sparse regression. The resulting SISSO-derived empirical expression for formation energy is

\begin{equation}
E_f^{SISSO} \approx c_0 + \sum_{i=1}^{4} c_i d_i,
\label{eq:formation}
\end{equation}

where $c_0 = 1.00$, $c_1 = -7.14$, $c_2 = 15.88$, $c_3 = 1.79$ and $c_4 = -0.01$. The nonlinear descriptors are defined as

\begin{equation}
d_1 = \frac{f_1}{f_2 f_5}, \qquad
d_2 = \exp\!\left(\frac{f_4}{f_3}\right), \qquad
d_3 = \frac{f_1 f_3}{f_4}, \qquad
d_4 = \frac{f_5}{\sqrt{f_6}},
\end{equation}

where $f_1$, $f_2$, $f_3$, $f_4$, $f_5$ and $f_6$ denote the normalized values for the electron affinity of the X-site element, the first ionization energy of the B$_1$-site element, the LUMO energy of the A-site molecule, the polarizability of the A-site molecule, the Mendeleev number of the B$_2$-site element and the Mendeleev number of the B$_1$-site element, respectively.

The SISSO expression in Eq.~(\ref{eq:formation}) describes formation energy through coupled combinations of molecular and elemental properties. The first descriptor, $d_1$, combines the electron affinity of the X-site element with the first ionization energy of the B$_1$-site element and the Mendeleev number of the B$_2$-site element. Its negative coefficient, $c_1$, indicates that increasing this ratio lowers the predicted formation energy when the other descriptors are held fixed, corresponding to greater energetic favourability relative to the reference states. Similarly, the negative coefficient $c_4$ indicates that increasing $d_4$, the ratio of the B$_2$-site Mendeleev number to the square root of the B$_1$-site Mendeleev number, favours a lower predicted formation energy under the same condition. The second and third descriptors ($d_2$ and $d_3$) couple the LUMO energy and polarizability of the A-site molecule, with the third additionally incorporating the X-site electron affinity. These terms link the molecular electronic properties to those of the inorganic constituents, highlighting their combined contributions to the energetic trends captured by the model.

% The results of the SISSO calculations presented multiple potential formulas to explain the bandgaps. These formulas were then used on the 180k new candidates dataset, before a correlation analysis was done in relation to the XGBoost predictions. Using this metric, the best performing formula had a Spearman correlation of 0.5207, and a Pearson correlation of 0.5384. The formula is 

% \begin{equation}
%     E_g\approx c_1\frac{X/averB}{e^{X-averB}} + c_2 \mathcal{I}_{B_1}/\sqrt{\mathcal{V}_{B_1}} + c_3(\mathcal{I}_{B_1}-\mathcal{I}_{B_2})*\mathcal{T}_{B_2} + c_4 
% \end{equation}

% where the constants $c_i$ are given by $c_1 = 7.5012410980$, $c_2 = -1.2164501360$, $c_3 = 2.425442731\times{10^{-4}}$, and $c_4 = -1.6921790460$. $\mathcal{I}$, $\mathcal{V}$ and $\mathcal{T}$ denote the first ionization energy, volume and melting point for the elemental crystal, respsectively.
% % [ADD A LITTLE MORE]

For bandgaps of perovskites, we performed SISSO on a dataset of 923 perovskite samples, with a maximum descriptor dimension of four. The selected model expresses bandgap as a linear combination of three nonlinear descriptors, together with an intercept, constructed from the eight primary features retained after SHAP ranking and correlation-based pruning. At each descriptor dimension, the SIS step retained the 100 highest-ranked candidate features (nf$\_$sis = 100) for subsequent sparse regression. The resulting SISSO-derived empirical expression is

\begin{equation}
E_{\mathrm{g}}^{SISSO} \approx a_0 + \sum_{i=1}^{4} a_i d_i,
\label{eq:bandgaps}
\end{equation}

where $a_0 = 3.70$, $a_1 = -1.85 \times 10^{-4}$, $a_2 = -0.01$, $a_3 = 3.89\times 10^{-3}$ and $a_4 = -1.15$. The nonlinear descriptors are defined as

\begin{equation}
d_1 = g_3 g_6, \qquad
d_2 = g_2 g_5 g_7, \qquad
d_3 = \sqrt{g_3}g_1, \qquad
d_4 = \frac{g_1}{g_7}.
\end{equation}

where $g_1$ denotes the normalized values of the difference between the atomic radius of the X-site element and the average atomic radius of the B$_1$- and B$_2$-site elements, corresponding to X-averB. $g_2$ denotes the normalized values of the quotient of the X-site element and the average atomic radius of the B$_1$ and B$_2$-site elements. $g_3$ denotes the normalized values of the first ionization energy of the B$_1$-site element. $g_4$ represents the normalized values of the radii of the B$_1$ site elements, and $g_5$ denotes the normalized values of the first ionization energy of the B$_2$ site elements; and $g_6$ denotes the melting temperature of the B$_2$ site element in its crystalline form. 

%X-averB B2_melting_point        B1_Firstionization      B1_atomic_volume        B1_melting_point        B1_mendeleev_number     B2_Firstionization

%The SISSO expression in Eq.~(\ref{eq:bandgaps}) describes band gap through coupled combinations of geometric and elemental properties. The first descriptor, $d_1$, combines the ratio of the X-site atomic radius to the average B-site atomic radius with an exponential function of their radius difference. Its positive coefficient, $a_1$, indicates that increasing this descriptor widens the predicted band gap when the other descriptors are held fixed, capturing an empirical dependence on the relative atomic sizes of the inorganic constituents. The second descriptor, $d_2$, combines the first ionization energy of the B$_1$-site element with the square root of its elemental atomic volume. Its negative coefficient, $a_2$, indicates that increasing this ratio lowers the predicted band gap under the same condition. The third descriptor, $d_3$, couples the first-ionization-energy difference between the B$_1$- and B$_2$-site elements to the melting temperature of the B$_2$-site element. Because $a_3$ is positive, this term contributes positively to the predicted band gap when the B$_1$-site ionization energy exceeds that of the B$_2$-site element, and negatively when the ordering is reversed. Together, these descriptors provide an interpretable representation of the compositional dependencies captured by the model, although they do not directly quantify orbital interactions or structural distortions.

Similarly, the SISSO expression in Eq.~(\ref{eq:bandgaps}) describes bandgap through coupled combinations of geometric and elemental properties as well. The first descriptor, $d_1$, combines the first ionization energy of the B$_1$-site element with its Mendeleev number. Its negative coefficient, $a_1$, indicates that increasing this product lowers the predicted bandgap when the other descriptors are held fixed, capturing an empirical dependence on B-site elemental identity and electronic character. The second descriptor, $d_2$, couples the melting temperatures of the B$_1$- and B$_2$-site elements in their elemental crystalline forms with $g_7$. Its negative coefficient, $a_2$, likewise associates an increase in this product with a lower predicted bandgap under the same condition. The third and fourth descriptors incorporate the difference between the X-site atomic radius and the average B-site atomic radius, weighted by the square root of the B$_1$-site first ionization energy and the inverse of $g_7$, respectively. Although $a_3$ is positive and $a_4$ is negative, the net dependence on the radius difference is governed by $a_3\sqrt{g_3}+a_4/g_7$, rather than either coefficient alone. Together, these descriptors provide an interpretable representation of the compositional dependencies captured by the model, without directly quantifying orbital interactions or structural distortions.

% When isolating for specific B-site elements, 5 elements showed Pearson and Spearman correlation above 0.4 between XGBoost and the SISSO function. Those elements were Cs, K, Li, Na, and Rb. The SISSO predictions for all 5 elements show very similar normal distributions, especially Cs, K, and Rb. 

% \begin{figure}[ht!]
%     \centering
%     \includegraphics[width=1\textwidth]{SISSO-new-vs-XGB-official.png}
%     \caption{SISSO.a) total prediction dataset, (b) Cs, (c) K, (d) Li, (e) Na, and (f)Rb sub-datasets, respectively.}
%     \label{fig:sisso}
% \end{figure}

\section{Conclusion}
In this work, we developed a multistage framework combining XGBoost, CGCNN, UPET and SISSO to screen 180,000 hybrid organic–inorganic perovskites. XGBoost achieved $R^2$ values of 0.99 for formation energy and 0.93 for bandgap. SHAP analysis identified the dominant molecular, elemental and geometric features, while correlation analysis removed redundant inputs. SISSO then derived compact analytical expressions from the retained features. The formation-energy SISSO-induced formulae captures coupled contributions from the organic molecule and inorganic framework, whereas the  SISSO-induced formulae emphasizes relative atomic sizes and B-site electronic properties, enabling efficient and interpretable materials discovery and screening.

\section*{Acknowledgements}
The authors acknowledge the use of ChatGPT (OpenAI) for assistance in improving the language and readability of the manuscript.

\section*{Competing Interests statement}
The authors declare no competing interests.

\section*{Data Availability}
The datasets generated during and/or analysed during the current study are available from the corresponding author on reasonable request.

\section*{Author Contributions}
Q. J. conceived the project and supervised the research. C. X. developed and trained the models and analyzed the results. C. X. wrote the original manuscript. C. X. and H. Z. contributed to the discussion of the results.

%\newpage
%\bibliographystyle{unsrt}
%\bibliography{references}

\end{document}